\documentclass[sigconf]{acmart}
\AtBeginDocument{%
  \providecommand\BibTeX{{%
    \normalfont B\kern-0.5em{\scshape i\kern-0.25em b}\kern-0.8em\TeX}}}

\usepackage[inline]{enumitem}
\usepackage{pbox}
\usepackage{balance}
\usepackage{todonotes}
\newlist{inlinelist}{enumerate*}{1}
\setlist*[inlinelist,1]{label=\roman*),itemjoin={{, }},itemjoin*={{, and }}}
\usepackage{url}
\usepackage[normalem]{ulem}

\begin{document}
\fancyhead{} 
\title{SPLADE: Sparse Lexical and Expansion Model \\ for First Stage Ranking}


\author{Thibault Formal}
\affiliation{%
  \institution{Naver Labs Europe}
  \city{Meylan}
  \country{France}
 }
 \affiliation{%
  \institution{Sorbonne Université, LIP6}
  \city{Paris}
  \country{France}
  }
  
\email{thibault.formal@naverlabs.com}

\author{Benjamin Piwowarski}
\affiliation{%
  \institution{Sorbonne Université, CNRS, LIP6}
  \city{Paris}
  \country{France}}
\email{benjamin.piwowarski@lip6.fr}

\author{Stéphane Clinchant}
\affiliation{%
  \institution{Naver Labs Europe}
  \city{Meylan}
  \country{France}}
\email{stephane.clinchant@naverlabs.com}





\begin{abstract}
  
In neural Information Retrieval, ongoing research is directed towards improving the first retriever in ranking pipelines. Learning dense embeddings to conduct retrieval using efficient approximate nearest neighbors methods has proven to work well. Meanwhile, there has been a growing interest in learning \emph{sparse} representations for documents and queries, that could inherit from the desirable properties of bag-of-words models such as the exact matching of terms and the efficiency of inverted indexes. In this work, we present a new first-stage ranker based on explicit sparsity regularization and a log-saturation effect on term weights, leading to  highly sparse representations and competitive results with respect to state-of-the-art dense and sparse methods. Our approach is simple, trained end-to-end in a single stage. We also explore the trade-off between effectiveness and efficiency, by controlling the contribution of the sparsity regularization. 

\end{abstract}






\keywords{neural networks, indexing, sparse representations, regularization}

\maketitle

\section{Introduction}

The release of large pre-trained language models like BERT~\cite{bert} has shaken-up Natural Language Processing and Information Retrieval.
These models have shown a strong ability to adapt to various tasks by simple fine-tuning. At the beginning of 2019, \textit{Nogueira and Cho}~\cite{passage_ranking} achieved state-of-the-art results -- by a large margin -- on the MS MARCO passage re-ranking task, paving the way for LM-based neural ranking models. Because of strict efficiency requirements, these models have initially been used as re-rankers in a two-stage ranking pipeline, where first-stage retrieval -- or candidate generation -- is conducted with bag-of-words models (e.g. BM25) that rely on inverted indexes. While BOW models remain strong baselines~\cite{neural_hype}, they suffer from the long standing vocabulary mismatch problem, where relevant documents might not contain terms that appear in the query. Thus, there have been attempts to substitute standard BOW approaches by learned (neural) rankers. Designing such models poses several challenges regarding efficiency and scalability: therefore there is a need for methods where most of the computation can be done offline and online inference is fast. 
Dense retrieval with approximate nearest neighbors search has shown impressive results \cite{xiong2021approximate,lin2020distill,ding2020rocketqa}, but is still combined with BOW models because of its inability to explicitly model term matching. Hence, there has recently been a growing interest in learning \emph{sparse representations} for queries and documents~\cite{snrm, dai2019contextaware,nogueira2019document,zhao2020sparta,sparterm2020}. By doing so, models can inherit from the desirable properties of BOW models like exact-match of (possibly latent) terms, efficiency of inverted indexes and interpretability. Additionally, by modeling implicit or explicit (latent, contextualized) \emph{expansion} mechanisms -- similarly to standard expansion models in IR -- these models can reduce the vocabulary mismatch. 

The contributions of this paper are threefold:
 \begin{enumerate*}
     \item we build upon SparTerm~\cite{sparterm2020}, and show that a mild tuning of hyperparameters brings improvements that largely outperform the results reported in the original paper;
    \item we propose the SParse Lexical AnD Expansion (SPLADE) model, based on a logarithmic activation and sparse regularization. SPLADE performs an efficient document expansion~\cite{sparterm2020, MacAvaney_2020}, with competitive results with respect to complex training pipelines for dense models like ANCE~\cite{xiong2021approximate};
     \item finally, we show how the sparsity regularization can be controlled to influence the trade-off between efficiency (in terms of the number of floating-point operations) and effectiveness.
 \end{enumerate*}



\section{Related Works}
Dense retrieval based on BERT Siamese models~\cite{sentence_bert} has become the standard approach for candidate generation in Question Answering and IR~\cite{guu2020realm,karpukhin2020dense,xiong2020approximate,lin2020distill,ding2020rocketqa}. 
While the backbone of these models remains the same, 
recent works highlight the critical aspects of the training strategy to obtain state-of-the-art results, ranging from improved negative sampling~\cite{xiong2020approximate,ding2020rocketqa} to distillation~\cite{hofstatter2020improving,lin2020distill}. ColBERT~\cite{colbert} pushes things further: the postponed token-level interactions allow to efficiently apply the model for first-stage retrieval, benefiting of the effectiveness of modeling fine-grained interactions, at the cost of storing embeddings for each (sub)term -- raising concerns about the actual scalability of the approach for large collections. 
To the best of our knowledge, very few studies have discussed the impact of using \emph{approximate} nearest neighbors (ANN) search on IR metrics~\cite{boytsov2018efficient, tu2020approximate}.
Due to the moderate size of the MS MARCO collection, results are usually reported  with an \emph{exact}, brute-force search, therefore giving no indication on the effective computing cost.

An alternative to dense indexes is term-based ones.
Building on standard BOW models, \textit{Zamani et al.} first introduced SNRM~\cite{snrm}: the model embeds documents and queries in a sparse high-dimensional latent space by means of $\ell_1$ regularization on representations.
However, SNRM effectiveness remains limited and its efficiency has been questioned~\cite{paria2020minimizing}. 
More recently, there have been attempts to transfer the knowledge from pre-trained LM to sparse approaches. 
Based on BERT, DeepCT~\cite{dai2019contextaware, 10.1145/3366423.3380258, 10.1145/3397271.3401204} focused on learning contextualized term weights in the full vocabulary space -- akin to BOW term weights. 
However, as the vocabulary associated with a document remains the same, this type of approach does not solve the vocabulary mismatch, as acknowledged by the use of query expansion for retrieval \cite{dai2019contextaware}.
A first solution to this problem consists in expanding documents using generative approaches such as doc2query~\cite{nogueira2019document} and docTTTTTquery~\cite{doct5} to predict expansion words for documents. The document expansion adds new terms to documents -- hence fighting the vocabulary mismatch -- as well as repeats existing terms, implicitly performing re-weighting by boosting important terms. These methods are however limited by the way they are trained (predicting queries), which is indirect in nature and limit their progress.
A second solution to this problem, that has been chosen by recent works such as~\cite{sparterm2020,MacAvaney_2020,zhao2020sparta}, is to estimate the importance of each term of the vocabulary \emph{implied by} each term of the document, i.e. to compute an interaction matrix between the document or query tokens and all the tokens from the vocabulary. This is followed by an aggregation mechanism (roughly sum for SparTerm~\cite{sparterm2020}, max for EPIC~\cite{MacAvaney_2020} and SPARTA~\cite{zhao2020sparta}), that allows to compute an importance weight for each term of the vocabulary, for the full document or query. 
However, EPIC and SPARTA (document) representations are not sparse enough by construction -- unless resorting on top-$k$ pooling -- contrary to SparTerm, for which fast retrieval is thus possible. Furthermore, the latter does not include (like SNRM) an \emph{explicit} sparsity regularization, which hinders its performance. Our SPLADE model relies on such regularization, as well as other key changes, that boost both the efficiency and the effectiveness of this type of models.

\section{Sparse Lexical representations for first-stage ranking}


In this section, we first describe in details the SparTerm model~\cite{sparterm2020}, before presenting our model named SPLADE.

\subsection{SparTerm}

SparTerm  predicts term importance -- in BERT WordPiece vocabulary ($|V|=30522$) -- based on the logits of the Masked Language Model (MLM) layer. More precisely, let us consider an input query or document sequence (after WordPiece tokenization) $t=(t_1,t_2,...,t_N)$, and its corresponding BERT embeddings $(h_1,h_2,...,h_N)$. We consider the importance $w_{ij}$ of the token $j$ (vocabulary) for a token $i$ (of the input sequence): 
\begin{equation}
    w_{ij} = \text{transform}(h_i)^T E_j + b_j \quad j \in \{1,...,|V|\}
    \label{equation_1}
\end{equation} 
where $E_j$ denotes the BERT input embedding for token $j$, $b_j$ is a token-level bias, and transform$(.)$ is a linear layer with GeLU activation and LayerNorm. Note that Eq.~\ref{equation_1} is equivalent to the MLM prediction, thus it can be also be initialized from a pre-trained MLM model. The final representation is then obtained by summing importance predictors over the input sequence tokens, after applying ReLU to ensure the positivity of term weights: 
\begin{equation}
w_j=g_j \times \sum_{i \in t}\label{eq_model} \text{ReLU}(w_{ij})
\end{equation}
where $g_j$ is a binary mask (gating) described latter.
The above equation can be seen as a form of query/document \emph{expansion}, as observed in \cite{sparterm2020,MacAvaney_2020}, since for each token of the vocabulary the model predicts a new weight $w_j$. 
%
SparTerm~\cite{sparterm2020} introduces two sparsification schemes that turn off a large amount of dimensions in query and document representations, allowing to efficiently retrieve from an inverted index: 

 { \bf lexical-only} is a BOW masking, i.e. $g_j=1$ if token $j$ appears in $t$, and 0 otherwise;

{\bf expansion-aware} is a lexical/expansion-aware binary gating mechanism, where $g_{j}$ is \emph{learned}. To preserve the original input, it is forced to 1 if the token $j$ appears in $t$.


\noindent Let $s(q,d)$ denote the ranking score obtained via dot product between $q$ and $d$ representations from Eq. \eqref{eq_model}. Given a query $q_i$, a positive document $d_i^+$ and a negative document $d_i^-$,  SparTerm is trained  by minimzing the following loss:

\begin{equation}\mathcal{L}_{rank} = - \log\frac{e^{s(q_i,d_i^+)}}{e^{s(q_i,d_i^+)} + e^{s(q_i,d_i^-)}}
\end{equation}


\paragraph{\bf Limitations}
SparTerm expansion-aware gating 
is somewhat intricate, and the model cannot be trained end-to-end: the gating mechanism is learned beforehand, 
and \emph{fixed} while fine-tuning the matching model with $\mathcal{L}_{rank}$, therefore preventing the model to learn the optimal sparsification strategy for the ranking task. Moreover, the two lexical and expansion-aware strategies do perform almost equally well, questioning the actual benefits of expansion.

\subsection{SPLADE: SParse Lexical AnD Expansion model}


In the following, we propose slight, but essential changes to the SparTerm model that dramatically improve its performance.

\paragraph{\bf Model} We introduce a minor change in the importance estimation from Eq.~\ref{eq_model}, by introducing a log-saturation effect which 
    prevents some terms to dominate and 
    naturally ensures sparsity in representations:
\begin{equation}
w_{j}=\sum_{i \in t} \log \left(1 + \text{ReLU}(w_{ij}) \right)
\label{eq_model_log} 
\end{equation}
While it is intuitive that using a log-saturation prevents some terms from dominating -- drawing a parallel with  axiomatic approaches in IR and $\log$(tf) models~\cite{10.1145/1008992.1009004} -- the implied sparsity can seem surprising at first, but,
according to our experiments, it obtains better experimental results and allows already to obtain sparse solutions \emph{without any regularization}. 


\paragraph{\bf Ranking loss} 

Given a query $q_i$ in a batch, a positive document $d_i^+$, a (hard) negative document $d_i^-$ (e.g. coming from BM25 sampling), and a set of negative documents in the batch (positive documents from other queries) $\{d_{i,j}^-\}_j$, we consider the ranking loss from~\cite{ding2020rocketqa}, which can be interpreted as the maximization of the probability of the document $d_i^+$ being relevant among the documents $d_i^+, d_i^-$ and $\{d_{i,j}^-\}$:

\begin{equation}\mathcal{L}_{rank-IBN} = - \log\frac{e^{s(q_i,d_i^+)}}{e^{s(q_i,d_i^+)} + e^{s(q_i,d_i^-)} + \sum_j e^{s(q_i,d_{i,j}^-)}}
\end{equation}

The \emph{in-batch negatives} (IBN) sampling strategy is widely used for training image retrieval models, and has shown to be effective in learning first-stage  rankers~\cite{karpukhin2020dense,ding2020rocketqa,lin2020distill}.

\paragraph{\bf Learning sparse representations}
The idea of learning sparse representations for first-stage retrieval dates back to SNRM~\cite{snrm}, via $\ell_1$ regularization. Later, \cite{paria2020minimizing} pointed-out that minimizing the $\ell_1$ norm of representations does not result in the most efficient index, as nothing ensures that posting lists are evenly distributed. Note that this is even more true for standard indexes due to the Zipfian nature of the term frequency distribution. 
To obtain a well-balanced index, \textit{Paria et al.} \cite{paria2020minimizing} introduce the \texttt{FLOPS} regularizer, a smooth relaxation of the average number of floating-point operations necessary to compute the score of a document, and hence directly related to the retrieval time. It is defined using $a_j$ as a continuous relaxation of the activation (i.e. the term has a non zero weight) probability $p_j$ for token $j$, and estimated for documents $d$ in a batch of size $N$ by
$\bar{a}_j=\frac{1}{N} \sum_{i=1}^N w^{(d_i)}_{j}$.
This gives the following regularization loss
$$
\ell_{\texttt{FLOPS}} = \sum_{j\in V} {\bar a}_j^2 = \sum_{j \in V} \left( \frac{1}{N} \sum_{i=1}^N  w_j^{(d_i)} \right)^2
$$
This differs from the $\ell_1$ regularization used in SNRM~\cite{snrm} where the ${\bar a}_j$ are not squared: using $\ell_{\texttt{FLOPS}}$ thus pushes down high average term weight values, giving rise to a more balanced index.

\paragraph{\bf Overall loss}
We propose to combine the best of both worlds for end-to-end training of sparse, expansion-aware representations of documents and queries. Thus, we discard the binary gating in SparTerm, and instead learn our log-saturated model (Eq. \ref{eq_model_log}) by jointly optimizing ranking and regularization losses: 
\begin{equation}
\mathcal{L} = \mathcal{L}_{rank-IBN} + \lambda_q \mathcal{L}^{q}_{\texttt{reg}} + \lambda_d \mathcal{L}^{d}_{\texttt{reg}}
\end{equation} 
where $\mathcal{L}_{\texttt{reg}}$ is a sparse regularization ($\ell_1$ or $\ell_{\texttt{FLOPS}}$). We use two distinct regularization weights ($\lambda_d$ and $\lambda_q$) for queries and documents -- allowing to put more pressure on the sparsity for queries, which is critical for fast retrieval.

\section{Experimental setting and results}

We trained and evaluated our models on the MS MARCO passage ranking dataset\footnote{\url{https://github.com/microsoft/MSMARCO-Passage-Ranking}} in the full ranking setting. The dataset contains approximately $8.8$M passages, and hundreds of thousands training queries with shallow annotation ($\approx 1.1$ relevant passages per query in average). The development set contains $6980$ queries with similar labels, while the TREC DL 2019 evaluation set provides fine-grained annotations from human assessors for a set of $43$ queries~\cite{craswell2020overview}. 

\paragraph{\bf Training, indexing and retrieval}

We initialized the models with the \texttt{BERT-base} checkpoint. Models are trained with the ADAM optimizer, using a learning rate of $2e^{-5}$ with linear scheduling and a warmup of $6000$ steps, and a batch size of $124$. We keep the best checkpoint using MRR@10 on a validation set of $500$ queries, after training for $150$k iterations (note that this is not optimal, as we validate on a re-ranking task). 
We consider a maximum length of $256$ for input sequences. In order to mitigate the contribution of the regularizer at the early stages of training, we follow \cite{paria2020minimizing} and use a scheduler for $\lambda$, quadratically increasing $\lambda$ at each training iteration, until a given step ($50$k in our case), from which it remains constant. Typical values for $\lambda$ fall between $1e^{-1}$ and $1e^{-4}$.
For storing the index, 
we use a custom implementation based on Python arrays, and we rely on Numba~\cite{lam2015numba} to parallelize retrieval. Models\footnote{We made the code public at \url{https://github.com/naver/splade}} are trained using PyTorch~\cite{paszke2019pytorch} and HuggingFace transformers~\cite{wolf2020huggingfaces}, on $4$ Tesla $V100$ GPUs with $32$GB memory. 

\paragraph{\bf Evaluation}
We report Recall@1000 for both datasets, as well as the official metrics MRR@10 and NDCG@10 for MS MARCO dev set and TREC DL 2019 respectively. Since we are essentially interested in the first retrieval step,  we do not consider re-rankers based on BERT, and we compare our approach to first stage rankers only -- results reported on the MS MARCO leaderboard are thus not comparable to the results presented here. We compare to the following sparse approaches
\begin{enumerate*}
    \item BM25
    \item DeepCT~\cite{dai2019contextaware} 
    \item doc2query-T5~\cite{doct5}
    \item and SparTerm~\cite{sparterm2020}
\end{enumerate*}, as well as state-of-the-art dense approaches ANCE~\cite{xiong2020approximate} and TCT-ColBERT~\cite{lin2020distill}. We report the results from the original papers. We include a pure lexical SparTerm trained with our ranking pipeline (ST lexical-only). To illustrate the benefits of the log-saturation, we add results for models trained using Eq. \eqref{eq_model} instead of Eq. \eqref{eq_model_log} (ST exp-$\ell_1$ and ST exp-$\ell_{\texttt{FLOPS}}$). 
For sparse models, we indicate an estimation of the average number of floating-point operations between a query and a document in Table \ref{table_1}, when available, which is defined as 
the expectation $\mathbb{E}_{q,d} \left[ \sum_{j \in V} p_j^{(q)}p_j^{(d)} \right]$
where $p_j$ is the activation probability for token $j$ in a document $d$ or a query $q$. It is empirically estimated from a set of approximately $100$k development queries, on the MS MARCO collection.


Results are given in Table \ref{table_1}.
Overall, we observe that:
\begin{enumerate*}
    \item \textit{our models outperform the other sparse retrieval methods by a large margin (except for recall@1000 on TREC DL);}
    \item \textit{the results are competitive with state-of-the-art dense retrieval methods.} 
\end{enumerate*}

More specifically, our training method for ST lexical-only already outperforms  the results of DeepCT as well as the results reported in the original SparTerm paper -- including the model using expansion. Thanks to the additional sparse expansion mechanism, we are able to obtain results on par with state-of-the-art dense approaches on MS MARCO dev set (e.g. Recall@1000 close to $0.96$ for ST exp-$\ell_1$), but with a much bigger average number of FLOPS. 

By adding a log-saturation effect to the expansion model, SPLADE greatly increases sparsity -- reducing the FLOPS to similar levels than BOW approaches -- at no cost on performance when compared to the best first-stage rankers. In addition, we observe the advantage of the FLOPS regularization over $\ell_1$ in order to decrease the computing cost. Note that in contrast to SparTerm, SPLADE is trained end-to end in a single step. It is also remarkably simple, compared to dense state-of-the-art baselines such as ANCE~\cite{xiong2020approximate}, and avoids resorting to approximate neighbors search, whose impact on IR metrics has not been fully evaluated yet.

\setlength{\tabcolsep}{2pt} 
\begin{table}
\small
\centering
\caption{Evaluation on MS MARCO passage retrieval (dev set) and TREC DL 2019}
\begin{tabular}{lccccc}
\toprule
model &  \multicolumn{2}{c}{MS MARCO dev} & \multicolumn{2}{c}{TREC DL 2019} & FLOPS \\
& MRR@10 & R@1000 & NDCG@10 & R@1000 & \\
\midrule
\texttt{Dense retrieval} & & & & \\
Siamese (ours) & 0.312 & 0.941 & 0.637 & 0.711 & -  \\
ANCE \cite{xiong2020approximate} & 0.330 & 0.959 & 0.648 & - & -  \\
TCT-ColBERT \cite{lin2020distill} & 0.335 & 0.964 & 0.670 & 0.720 & - \\

\bottomrule
\hline
\texttt{Sparse retrieval} & & & & &\\
BM25 & 0.184 & 0.853 & 0.506 & 0.745 & 0.13 \\ 
DeepCT \cite{dai2019contextaware} & 0.243 & 0.913 & 0.551 & 0.756 & - \\ 
doc2query-T5 \cite{doct5} & 0.277 & 0.947 & 0.642 & 0.827  & 0.81\\ 
ST lexical-only \cite{sparterm2020} & 0.275 & 0.912 & - & - & - \\ 
ST expansion \cite{sparterm2020} & 0.279 & 0.925 & - & -    & -\\ 
\bottomrule
\hline
\texttt{Our methods} & & & & & \\
ST lexical-only & 0.290 & 0.923 & 0.595 & 0.774 & 1.84 \\
ST exp-$\ell_1$  & 0.314 	& 0.959  &  0.668 & 0.800 & 4.62   \\
ST exp-$\ell_{\texttt{FLOPS}}$  & 0.312 &	0.954   & 0.671 & 0.813 & 2.83  \\
SPLADE-$\ell_1$  & 0.322 & 0.954 & 0.667 & 0.792 & 0.88 \\
SPLADE-$\ell_{\texttt{FLOPS}}$  & 0.322 & 0.955 & 0.665 & 0.813 & 0.73  \\
\end{tabular}
\label{table_1}
\end{table}



\begin{figure}[b]
  \caption{Performance vs FLOPS for SPLADE models trained with different regularization strength $\lambda$ on MS MARCO}
  \includegraphics[width=0.45\textwidth]{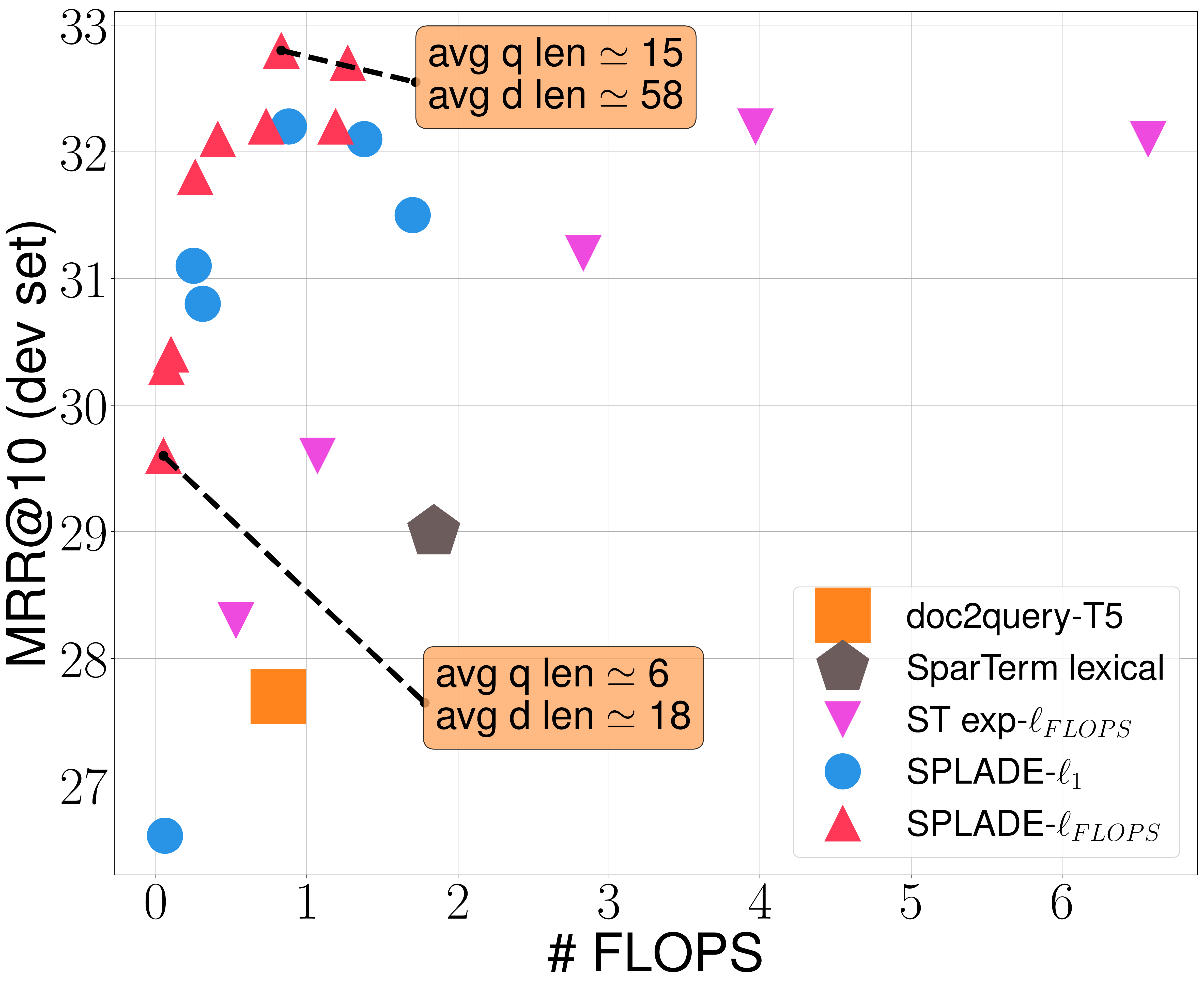}
  \label{perf_flops}
\end{figure}


\paragraph{\bf Effectiveness-efficiency trade-off}
Figure \ref{perf_flops} illustrates the trade-off between effectiveness (MRR@10) and efficiency (FLOPS), when we vary $\lambda_q$ and $\lambda_d$ (varying both implies that plots are not smooth). We observe that ST exp-$\ell_{\texttt{FLOPS}}$ falls far behind BOW models and SPLADE in terms of efficiency. In the meantime, SPLADE reaches efficiency levels equivalent to sparse BOW models, while outperforming doc2query-T5.
Interestingly, strongly regularized models still show competitive performance (e.g. \texttt{FLOPS}=$0.05,$ MRR@10=$0.296$). Finally, the regularization effect brought by $\ell_{\texttt{FLOPS}}$ compared to $\ell_1$ is clear: for the same level of efficiency, performance of the latter is always lower. 

\paragraph{\bf The role of expansion}

Experiments show that the expansion brings improvements w.r.t. to the purely lexical approach by increasing recall. Additionally, representations obtained from expansion-regularized models are sparser: the models learn how to balance expansion and compression, by both turning-off irrelevant dimensions and activating useful ones. On a set of $10$k documents, the SPLADE-$\ell_{\texttt{FLOPS}}$ from Table \ref{table_1} drops in average $20$ terms per document, while adding $32$ expansion terms. For one of our most efficient model (\texttt{FLOPS}=$0.05$), $34$ terms are dropped in average, for only $5$ new expansion terms. In this case, representations are extremely sparse: documents and queries contain in average $18$ and $6$ non-zero values respectively, and we need less that $1.4$ GB to store the index on disk.
Table \ref{tab:example} shows an example where the model performs term re-weighting by emphasizing on \emph{important} terms and discarding most of the terms without information content. Expansion allows to enrich documents, either by implicitly adding stemming effects (legs $\rightarrow$ leg) or by adding relevant topic words (e.g. treatment).

\begin{table}
\small
\centering
\caption{Document and expansion terms: { \small between parenthesis is the weight associated with the term -- omitted for the second occurrence of the term in the document, and strike-through for zeros} }
\begin{tabular}{p{\columnwidth}}
\toprule
    \multicolumn{1}{c}{\textbf{original document (doc ID: 7131647)}}\\
\midrule
{if (1.2) bow (2.56) legs (1.18) \sout{is} caused (1.29) by (0.47) \sout{the} bone (1.2) alignment (1.88) issue (0.87) \sout{than you may be} able (0.29) \sout{to} correct (1.37) through (0.43) \emph{bow legs} correction (1.05) \sout{exercises}. \sout{read more here..} \emph{if bow legs is caused by the bone alignment issue than you may be able to correct through bow legs correction exercises.}} \\
\midrule
    \multicolumn{1}{c}{\textbf{expansion terms}}\\
\midrule
(leg, 1.62) (arrow, 0.7) (exercise, 0.64) (bones, 0.63) (problem, 0.41) (treatment, 0.35) (happen, 0.29) (create, 0.22) (can, 0.14) (worse, 0.14) (effect, 0.08) (teeth, 0.06) (remove, 0.03) \\
\bottomrule
\hline
\end{tabular}
\label{tab:example}
\end{table}

\section{Conclusion}
Recently, dense retrieval based on BERT has demonstrated its superiority for first-stage retrieval, questioning the competitiveness of traditional sparse models. In this work, we have proposed SPLADE, a sparse model revisiting query/document expansion. Our approach relies on in-batch negatives, logarithmic activation and FLOPS regularization to learn effective and efficient sparse representations. SPLADE is an appealing candidate for initial retrieval: it rivals the latest state-of-the-art dense retrieval models, its training procedure is straightforward, its sparsity/FLOPS can be controlled explicitly through the regularization, and it can operate on inverted indexes. In reason of its simplicity, SPLADE is a solid basis for further improvements in this line of research.




\bibliographystyle{ACM-Reference-Format}
\balance
\bibliography{sample-base}

\end{document}